\documentclass[aps,pre,twocolumn,groupedaddress,showpacs]{revtex4}
\usepackage{epsf}
\newcommand{\be}{\begin{equation}}
\newcommand{\ee}{\end{equation}}
\newcommand{\ba}{\begin{eqnarray}}
\newcommand{\ea}{\end{eqnarray}}
\newcommand{\baa}{\begin{eqnarray}}
\newcommand{\eaa}{\end{eqnarray}}
\newcommand{\ed}{\end{document}}

\renewcommand{\baselinestretch}{1.2}
\begin{document}
\title{The $\phi\to \omega \pi^0 $ Decay in  the Chiral Model }
\author {K.R. Nasriddinov, B.N.Kuranov U.A. Khalikov\\
 and N.E.Iskandarov.}
\affiliation{ Uzbekistan,Tashkent-700100,Yusuf Khos Khojib-103.
Tashkent State Pedagogical University named after Nizami}
\begin{abstract}
The  $\phi \to \omega \pi^0$  decay is studied using the method of phenomenological
chiral Lagrangians. It is shown that the weak hadronic current between $\omega$-
and $\pi^0$ -mesons is equal to zero and this decay channel proceeds only due
to the $\phi-\rho$ mixing diagram.\hfill\break
\end{abstract}
\pacs{13.25.Jx, 12.39.Fe}
\maketitle

Recently, experimental evidence \cite{par1} of the G-parity violating
$\phi\to\omega\pi^0$ decay with the partial widths
$$
B(\phi\to\omega \pi^0)=(4.8^{+1.9}_{-1.7}\pm 0.8)*10^{-5}
$$
has been seen. At present the world average [2] for this decay channel is estimated to be
$$
 B(\phi \to \omega \pi^{0})=(5.2^{+1.3}_{-1.1})*10^{-5}.
$$

Here, we consider the $\phi\to\omega\pi^{0}$  decay on the basis of the method
of phenomenological chiral Lagrangians(PCL's)\cite{nee}. Studies of this decay channel
is of interest in this model for the following reasons:\\
First, this decay channel is a unique "laboratory" for verification of weak
hadron currents between pseudoscalar and vector meson states which was obtained
in\cite{kar}  within the formalism of phenomenological chiral Lagrangians
\begin{equation}
\label{}
I^{i}_{\mu}=g F_{\pi}\vartheta_{\mu}^{a}\varphi^{b} f_{abi},
\end{equation}
$g$ is the "universal" coupling constant fixed from the experimental
$\rho\to\pi\pi$ decay widht $g^{2}/4\pi=3.2$, $F_{\pi}=93 MeV$, $\vartheta^{a}_{\mu}$
and $\varphi^{b}$ are the fields of the $1^{-}$ and $0^{-}$ mesons, respectively,
and $f_{abi}$ are the structure constants of the $SU(3)$ group (a,b ,i = 1,...,8).\\
Second, this decay allows to study the nature of $\rho,\omega$ and $\phi-$  meson mixing.
Note,that in references \cite{nut,nut1}  the problems of $\pi^{0}-\eta$ - and $\omega-\phi$
mixings have been studied on the basis of this model as well and obtained
reasonable results for the $\tau$-lepton decay probabilities. In this calculation,
we used $\omega-\phi$-mixing [2] in the form
$$ \phi=\omega_{8} cos\theta_{V}-\omega_{1}sin\theta_{V}$$
$$\omega=\omega_{8} sin\theta_{V}+\omega_{1}cos\theta_{V}$$

According to the  expression for weak hadronic currents between pseudoscalar and
vector meson states (1), the Born amplitude of this decay is equal to zero,
because of structure constants $f_{13i}=f_{83i}=0,$ (in other words, the current
$I_{\mu}^{i}$ responsible for the direct $\phi\to\omega\pi^{0}$ decay (FIG.1) is zero).

According to the method of phenomenological chiral Lagrangians(PCL's), this decay channel
would originate via the intermediate $\bar D^{0*}$ $-$ meson state
(FIG. 2). In this case the weak interaction  Lagrangian  between $\phi$ and
$\bar D^{0*}$ mesons has the form given in reference \cite{nul7}
$$
\L_{W}^{(\Delta C=1)}=(1/2)^{-1/2} G_{F} h_{5} (-\sqrt{3} I_{\mu}^{8}) I_{\mu}^{9-i10}+H.c.,
$$
where ${G_{F}=10^{-5}}/{m_{p}^{2}}$ is the Fermi constant, $h_{5}$=0.285 is the factor
that describe deviations from the 20$-$plet dominance and which is determined
by the angles of current rotation about $7^{th}$ and $10^{th}$ axes in the SU(4) space, and
$$
I_{\mu}^{8}=(\sqrt{2}/g_{\rho}) m_{\vartheta}^{2} \vartheta_{\mu}^{8}=(\sqrt{2}/g_{\rho}) m_{\phi}^{2}\phi_{\mu} cos\theta_{V}.
$$
$$
I_{\mu}^{9-i10}=(\sqrt{2}/g_{\rho}) m_{D}^2 \bar D_{\mu}^{0*}.
$$
In the PCL the strong interaction Lagrangian of vector mesons with vector and pseudoscalar mesons has the form
\begin{equation}
\label{}
L_{S}(vv\varphi)=-1/4 g_{vv\varphi}\varepsilon_{\mu va\beta}(d_{kln}+if_{kln})(\partial_{\mu} V_{v}^{k}\partial_{\alpha}V_{\beta}^{l}\varphi^{n}),
\end{equation}
where $g_{vv\varphi}=(3 g^{2})/16\pi^{2} F_{\pi}$ is the coupling constant;
$d_{kln}$ and $f_{kln}$ are the symmetric and anti-symmetric structure constants of the SU(3) group, respectively.
Taking into account
$$
\bar D^{0*}=(1/2i)^{-1/2}(\vartheta_{9}-i\vartheta_{10}),
$$
$$
\pi^{0}=\varphi^{3},
$$
it should be noted, that  all structure constants responsible for this interaction equal zero
$$
f_{391} = f_{3101} = f_{398}=f_{3108}=d_{391}=d_{3101}=$$
$$
=d_{398}=d_{3108}=0.
$$
Therefore, the second  diagram does not give any contribution to
$\phi\to\omega\pi^{0}$ decay.

The next diagram (FIG. 3) also does not contribute to the partial width for the
$\phi\to\omega\pi^{0}$ decay. In this case  the Lagrangian of the strong coupling
of axial-vector mesons to vector and pseudoscalar mesons is derived in a similar
way and has the form \cite{nut}

\begin{equation}
\label{}
\L_{S}(0^{-},1^{+},1^{-})=-F_{\pi} g^{2}\varphi^{l} a_{\mu}^{i}\vartheta_{\mu}^{k} f_{lki},
\end{equation}
where $a_{\mu}^{i}$ are the fields of $1^{+}$ mesons.

The structure constants of the SU(3) group responsible for this transition
are equal zero $f_{391}=f_{3101}=f_{398}=f_{3108}=0$. It should be noted that the
diagrams 2 and 3 do not contribute to the partial width of the $\phi\to\omega\pi^{0}$
decay channel which is obvious also due to the hadronic flavor conservation principle.
According to the expression (2),also the anomalous diagram (FIG.4.) does not
contribute to the partial width of the $\phi\to\omega\pi^{0}$ decay because
$$
d_{311}=d_{388}=0.
$$

Finally the diagram with the intermediate $\omega$ meson (FIG. 5) does not contribute to the
partial width of this decay channel also because of these structure constants.

Within the method of phenomenological chiral Lagrangians(PCL's), the partial width of
the $\phi\to\omega\pi^{0}$ decay is therefore given by the diagrams with $\phi-\rho$ and $\omega-\rho$
mixings (FIG.6 and FIG.7). In this case all the structure constants are equal to zero except
$$
d_{331}=(1/2)^{-1/2},  d_{338}=(1/3)^{-1/2}.
$$

Note, that in [5] we studied the $\tau^{-}\to\pi^{-}\eta\nu_{\tau}$ decay of
the $\tau$ lepton in the framework of this method with taking into account
isotopic violation of chiral symmetry in the Oakes scheme.In this case the
$\pi^{0}-\eta$ - mixing Lagrangian has the form
\begin{equation}
\label{}
L_{\pi^{0}/\eta}=(-1/3)^{-1/2} m_{\pi^{0}}^{2}\pi^{0}\eta.
\end{equation}
It was shown, that this decay channel  was suppressed with respect to the
$\tau^{-}\to\pi^{-}\pi^{0}\nu_{\tau} $ decay by a factor $10^{-4}$. Therefore, it is
natural that the $\phi\to\omega\pi^{0}$ decay is strongly suppressed because of
$\phi-\rho$ and $\omega-\rho$ mixings.

According to (2), the Lagrangians describing $\rho$ -meson interaction with $\omega$
and $\pi{^0}$ mesons (FIG. 6), and $\phi$-meson interaction with $\rho$ and $\pi^0$ mesons (FIG.7) have
the forms, respectively
$$
L_{s}(\rho\to\omega\pi{^0})=-\frac{1}{4}(\frac{1}{\sqrt{3}}sin{\theta_{V}}+\frac{1}{\sqrt{2}}cos{\theta_{V}}\times
$$
\begin{equation}
\label{}
g_{vv\varphi}\varepsilon_{\mu\nu\alpha\beta}(\partial_{\mu}\omega_{\nu}\partial_{\alpha}\rho_{\beta}\pi^{0})
\end{equation}
$$
L_{s}(\phi\to\rho\pi{^0})=-\frac{1}{4}(\frac{1}{\sqrt{2}}sin{\theta_{V}}-\frac{1}{\sqrt{3}}cos{\theta_{V}})
\times
$$
\begin{equation}
\label{}
g_{vv\varphi}\varepsilon_{\mu\nu\alpha\beta}(\partial_{\mu}\phi_{\nu}\partial_{\alpha}\rho_{\beta}\pi^{0})
\end{equation}
At $\theta_{V}=39^{0}$ the factors of these Lagrangians are equal,respectively
$$
\frac{1}{4}(\frac{1}{\sqrt{3}}sin{\theta_{V}}+\frac{1}{\sqrt{2}}cos{\theta_{V}})=0,23
$$
$$
\frac{1}{4}(\frac{1}{\sqrt{2}}sin{\theta_{V}}-\frac{1}{\sqrt{3}}cos{\theta_{V}})=0,0005
$$

Therefore, the contribution of the diagram with $\omega-\rho$ mixing (FIG.7) is negligible
with respect to the contribution of the diagram with $\phi-\rho$ mixing (FIG.6).
Thus,the decay width of the $\phi\to\omega\pi^{0}$ decay mainly is defined by the diagram with $\phi-\rho$ mixing (FIG. 6).

Here, we estimate the contribution of this diagram to the partial width
of the $\phi\to\omega\pi^{0}$ decay using the $\pi^0-\eta$ mixing Lagrangian (4)
by making the substitutions $\pi^{0}\to\rho$, $\eta\to\phi $
\begin{equation}
\label{}
L_{\phi\rho}=\frac{m_{\rho}^{2}}{\sqrt{3}}\phi\rho
\end{equation}
The decay rate is given
\begin{equation}
\label{}
\Gamma=\frac{1}{3}\frac{1}{2m}\left|{M}\right|^{2}\Phi,
\end{equation}
where $m$ is the $\phi$-meson mass. The amplitude of the $\phi\to\omega\pi^{0}$
decay will be determined in accordance with (5) and (7),and has the form
$$
M=\frac{1}{2\sqrt{3}}(\frac{1}{\sqrt{3}}sin\theta_{V}+\frac{1}{\sqrt{2}}cos\theta_{V})^{2}\times
$$
$$
g_{vv\varphi}\frac{m_{\rho}^{2}}{m^{2}-m_{\rho}^{2}}\left[m^{2}m_{\omega}^{2}-\left(\frac{m^{2}+m_{\omega}^{2}-m_{\pi}^{2}}{2}\right)^{2}\right]
$$
The phase space has the form
$$
\Phi=\frac{1}{8\pi m^{2}}\left[m^{2}-(m_{\omega}+m_\pi)^{2}\right]^{1/2}[m^{2}-(m_{\omega}-m_\pi)^{2}]^{1/2}
$$
where $m_{\omega}$, $m_{\pi}$ and $m_{\rho}$ are the masses of $\omega$, $\pi^{0}$ and $\rho$ mesons,
respectively. According to (8) the decay rate has the form

$$\Gamma=\frac{1}{4608\pi m^{3}}(\frac{1}{\sqrt{3}}sin\theta_{V}+\frac{1}{\sqrt{2}}cos\theta_{V})^{2}g_{vv\varphi}^{2}(\frac{m_{\rho}^{2}}{m^{2}-m_{\rho}^{2}})^{2}$$
\begin{equation}
\label{}
\times \left[m^{2}-(m_{\omega}+m_\pi)^{2}\right]^{3/2}\left[m^{2}-(m_{\omega}-m_\pi)^{2}\right]^{3/2}
\end{equation}
It follows that the decay rate of the $\phi\to\omega\pi$ decay is
$$
\Gamma(\phi\to\omega\pi)=0.27 MeV,
$$
here  we took into account only the contribution of the intermediate $\rho$(770) meson state.
The decay rate with taking into account the contributions of intermediate $\rho$(1450) and $\rho$(1700)
meson states is
$$
\Gamma(\phi\to\omega\pi)=0.77 MeV,
$$
For the partial decay rate we obtain
$$
B(\phi\to\omega\pi)=0.18,
$$
that is four orders of magnitude large than the experiment \cite{nul}.Therefore, our estimation of the $\phi\to\omega \pi^0$ decay rate using $\pi^0\to\eta$ mixing Lagrangian is inconsistent with experimental data and study of this process in the framework of the $\phi-\rho$ mixing is of interest.

In summary, we have studied the $\phi\to\omega\pi^{0}$ decay in terms of chiral
Lagrangians,and we have shown that there is no direct $\phi\to\omega\pi^{0}$
decay (as was indicated in reference\cite{par1}); therefore, this decay channel with
G-parity violation will originates mainly from  $\phi-\rho$ - mixing. Therefore, the
$\phi\to\omega\pi^{0}$ decay is a unique "laboratory" for studing of the
nature of the $\rho-\omega-\phi$ - mixing. At present the investigation of this mixing
in the framework of phenomenological chiral Lagrangians is in progress.
\begin{center}
\vskip 2 \baselineskip
\bf {ACKNOWLEDGMENTS}
\end{center}
\vskip 1 \baselineskip

We are grateful to Prof.s. F. Hussain, R. Riazuddin and E. Predazzi for interest in this study and for stimulating discussions. One of authors (KRN) would like to thank Prof. S. Randjbar-Daemi for the hospitality
at the Abdus Salam ICTP during the course of this work.

\newpage
\begin{figure}
\begin{center}
   \epsfysize=10cm
\epsffile{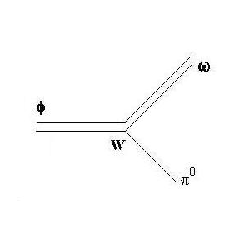}

   \vspace{.5cm}

\caption{\label{3}
The main diagram for the $\phi\to \omega \pi^0 $ decay, (W) weak-interaction vertex.}
\end{center}
\end{figure}

 \newpage
\begin{figure}
\begin{center}
   \epsfysize=10cm
\epsffile{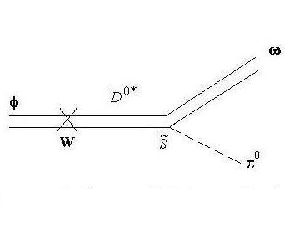}

   \vspace{.5cm}

\caption{\label{3}
The diagram with the intermediate $\bar D^{0*}$ $-$ meson,
$(\tilde S) $ anomalous strong-interection vertex}.
\end{center}
\end{figure}

\newpage
\begin{figure}
\begin{center}
   \epsfysize=10cm
\epsffile{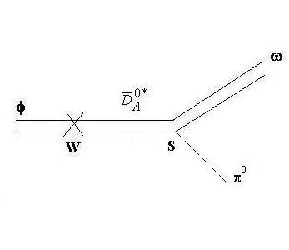}

   \vspace{.5cm}

\caption{\label{3}
The diagram with the intermediate $\bar D^{0*}_{A}$ - meson,
$(S)$ strong-interection vertex}.
\end{center}
\end{figure}

\newpage
\begin{figure}
\begin{center}
   \epsfysize=10cm
\epsffile{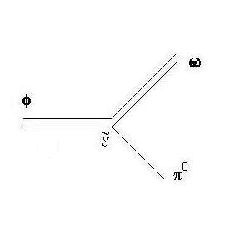}

   \vspace{.5cm}

\caption{\label{3}
The anomalous diagram for the $\phi\to \omega \pi^0 $ decay}
\end{center}
\end{figure}

\newpage
\begin{figure}
\begin{center}
   \epsfysize=10cm
\epsffile{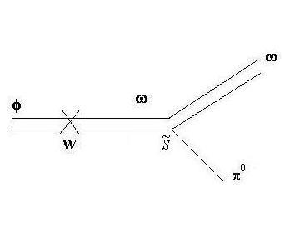}

   \vspace{.5cm}

\caption{\label{3}
 The diagram with the intermediate $\omega$ - meson.}
\end{center}
\end{figure}

\newpage
\begin{figure}
\begin{center}
   \epsfysize=10cm
\epsffile{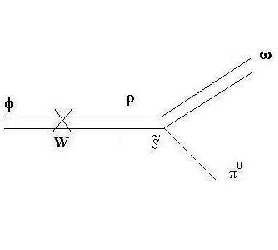}

   \vspace{.5cm}
\caption{\label{3}
The diagram with the $\rho-\phi$ - mixing.}
\end{center}
\end{figure}

\newpage
\begin{figure}
\begin{center}
   \epsfysize=10cm
\epsffile{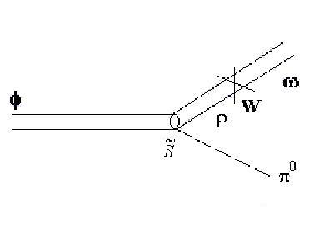}

   \vspace{.5cm}

\caption{\label{3}
The diagram with the $\rho-\omega$ mixing.}
\end{center}
\end{figure}

\end{document}